# A three-state kinetic mechanism for scaffold mediated signal transduction

Jason W. Locasale

Department of Biological Engineering, Massachusetts Institute of Technology,
77 Massachusetts Ave., Cambridge, MA 02139.
Locasale@MIT.edu

**Abstract**

Signaling events in eukaryotic cells are often guided by a scaffolding protein. Scaffold proteins assemble multiple proteins in a spatially localized signaling complex and exert numerous physical effects on signaling pathways. To study these effects, we consider a minimal, three-state kinetic model of scaffold mediated kinase activation. We first introduce and apply a path summation technique to obtain approximate solutions to a single molecule master equation that governs protein kinase activation. We then consider exact numerical solutions. We comment on when this approximation is appropriate and then use this analysis to illustrate the competition of processes occurring at many time scales involved in signal transduction in the presence of a scaffold protein. The findings are consistent with recent experiments and simulation data. Our results provide a framework and offer a mechanism for understanding how scaffold proteins can influence the shape of the waiting time distribution of kinase activation and effectively broaden the times over which protein kinases are activated in the course of cell signaling.

## Introduction

Cells detect external signals in the form of stresses, growth factors, DNA damage, hormones, among many others, and integrate them to achieve an appropriate biological response(1). Biochemical modifications in the form of reversible phosphorylations by enzymes known as kinases are detected by proteins to form networks that are used to integrate these signals(2). These complex networks are comprised of many modular structures that allow for many different biological responses. Signal propagation through these networks is often guided by scaffold protein(3). Scaffold proteins assemble multiple kinases (that are activated sequentially in a cascade) in close proximity to form controlled signaling complexes. Scaffold proteins are believed to regulate biochemical signaling pathways in a multitude of ways(3-5).

Experiments have suggested that the scaffold proteins have profound effects on regulating signaling dynamics(6-8). In particular, a key parameter is believed to be the concentration of scaffold proteins. Recent simulation results(9), which elaborated on these findings, showed that one effect that the concentration of scaffold proteins may have is to control the shape of the waiting time distribution of activation. As a result, scaffolds can then determine precisely the reactive flux of kinase activation that serves as a signal output to numerous downstream signaling pathways.

Here, we present a course grained, minimal, kinetic model that illustrates how the waiting time distribution of protein kinase activation is modified by the presence of different amounts of scaffold protein. The model involves multiple states in which a single protein kinase, situated at the end of a cascade, resides and corresponding transitions between these states are allowed(10). We analyze the resulting master equation by first introducing an approximate scheme that involves a weighted path summation over the possible trajectories that an individual kinase can take in the course of its transition from an inactive to an active state(11). We also consider exact numerical solutions. We find that, consistent with known simulation results, in certain limits the waiting time distribution of activation sharply decays and is effectively characterized by a single exponential whereas in other regimes, the waiting time distribution takes on a more complicated form. Our model provides a simple mechanistic description for how scaffold proteins and differences in their concentrations may regulate the waiting time distribution of kinase activation.

**Time scales for signal transduction via scaffold proteins**

Let us first consider physically, the time scales involved in scaffold mediated cell signaling. The signaling event that we consider consists of the sequential activation of multiple enzymes (kinases) in a cascade.

Consider the processes that must occur in order for a kinase at the end of a scaffolded kinase cascade to become activated. In this picture, we follow the trajectory a single molecule as it interacts with the scaffold and the upstream components of the pathway in which it is involved. Within the biochemical pathway, kinases in solution must encounter its targeted substrates by diffusion; therefore, encounter (or diffusion) times for the kinase in a sequence of a multi-tiered biochemical cascade, are important. These encounter times $\tau_{ec}$, determined by diffusive motion of proteins, behave as $\tau_{ec} \sim \left(D\rho^{d/2}\right)^{-1}$ in d dimensions, for a concentration, $\rho$ and diffusivity $D$. Other time scales ($\tau_k, \tau_p, \tau_{on}, \tau_{off}$) arise from rates of catalysis and protein-protein interactions such as the binding of a kinase to a scaffold. These times are for activation ($\tau_k$) and deactivation ($\tau_p$) by a kinase and phosphatase (an enzyme that removes a phosphate group) as well as for binding ($\tau_{on}$) and unbinding ($\tau_{off}$) to and from a scaffold.

We investigated the dynamical consequences of a model in which, in the course of activation of a single kinase, the collection of microscopic processes described above interacts with relative scaffold density $\zeta$ to give rise to several processes (involving state transitions of single molecule) with eight characteristic time scales $\tau_i$; $\tau_i \in \{\tau_1,...,\tau_8\}$. Scaffold density $\zeta$ has been shown in many contexts to be a key variable regulating signal transduction. If too few scaffolds are present, signaling occurs predominantly in solution. If too many scaffolds are present, proteins kinases exist predominantly in complexes that are incompletely assembled. There exists, therefore, an optimal concentration to assemble complete signaling complexes of multiple kinases(3). Schematics of these different scenarios are shown on the bottom of Fig. 1.

Kinases in solution could in principle, upon binding to a scaffold, be assembled into a complex that can not effectively signal. This is because the complex does not have a complete set of kinases bound to it. Such a kinase then would be trapped in a signaling incompetent state until it either disassociates from the complex or the requisite kinases upstream bind to the complex. Association and disassociation of kinases to and from incompletely assembled complexes are denoted by times $\tau_1$ and $\tau_2$ and are functions of scaffold density, diffusion times, catalysis rates, and binding kinetics.

This scenario also requires additional time scales: the times required for a kinase to bind and disassociate from solution into a signaling incompetent complex ($\tau_3$ and

$\tau_4$ respectively) and the times required for a signaling incompetent complex to transition to a signaling competent complex ($\tau_5$ and $\tau_6$ respectively) must be accounted for. Finally activation of the given kinase can occur with times $\tau_7$ and $\tau_8$ that involve diffusion and catalysis. The eight time scales comprising the model are functions of the following microscopic times:

$$\tau_1 = f(\tau_{ec}, \tau_{on}, \tau_{off}, \tau_k, \tau_p, \zeta)$$
$$\tau_2 = f(\tau_{off})$$
$$\tau_3 = f(\tau_{ec}, \tau_{on}, \tau_{off}, \tau_k, \tau_p, \zeta)$$
$$\tau_4 = f(\tau_{ec}, \tau_{on}, \tau_{off}, \tau_k, \tau_p, \zeta)$$
$$\tau_5 = f(\tau_{ec}, \tau_{on}, \tau_{off} \tau_k, \tau_p, \zeta)$$
$$\tau_6 = f(\tau_{off})$$
$$\tau_7 = f(\tau_{ec}, \tau_k, \tau_p)$$
$$\tau_8 = f(\tau_{ec}, \tau_k, \tau_p)$$

We investigated the consequences of such a minimal scenario that involves the dynamics of competing processes occurring at these eight phenomenological time scales.

**A Markov model illustrating the competition between many processes**

We considered this minimal course grained kinetic model in which a kinase, at the end of a biochemical cascade, $K_i$ can transition between four states denoted with four subscripts: in solution (S), bound to a signaling competent complex (C), bound to an signaling incompetent complex (I), and activated (A). Any bound kinase that is a part of an incomplete complex is said to be in state I. Fig. 1 shows a graph of the stochastic transitions to neighboring states that involve random waiting times that correspond to a set of eight random variables; Markovian dynamics are considered. The waiting time for a kinase to transition to a neighboring state is then Poisson distributed with time constants, $k_i; k_i \in \{k_1, k_2, ..., k_8\}$. Thus for the i[th] process, the waiting time distribution, $F(\tau_i)$, is the probability density for the first passage time distribution (FPT) and takes the form: $F(\tau_i) = k_i e^{-k_i \tau_i}$. Ultimately, the quantity of interest is the first-passage time distribution $F(t)$ (or its integrated value) for a kinase to transition to its activated state which we denote by $F(t)$. $F(t)$ is the time derivative of the cumulative probability distribution (CDF), $P(\tau_A < t) \equiv \int_0^t F(\tau')d\tau'$; so that $\frac{d}{dt}P(\tau_A < t) = F(t)$ where $\tau_A$ is the random waiting time for activation of a protein kinase. The survival probability $S(t)$ is related to the FPT and the CDF in the following way,

$$S(t) = \int_t^\infty F(t')dt' = 1 - \int_0^t F(t')dt' = 1 - P(\tau_A < t).$$ The kinetic equation, with an absorbing boundary condition at arrival at state $K_A$, is written as follows:

$$\frac{d}{dt}\vec{P} = Q \cdot \vec{P} \tag{1}$$

where $\vec{P}(t) = \begin{bmatrix} P_{K_A}(t) & P_{K_S}(t) & P_{K_C}(t) & P_{K_I}(t) \end{bmatrix}^T$ and

$$Q = \begin{pmatrix} 0 & k_7 & k_8 & 0 \\ 0 & -(k_1 + k_5 + k_7) & k_6 & k_2 \\ 0 & k_5 & -(k_4 + k_6 + k_8) & k_3 \\ 0 & k_1 & k_4 & -(k_2 + k_3) \end{pmatrix} \tag{2}$$

with the initial condition, $\vec{P}(0) = \begin{bmatrix} 0 & C_S & C_C & C_I \end{bmatrix}^T$ where $C_S$, $C_C$, and $C_I$ are the probabilities that a given kinase initially resides in the solution, complexed, or incompletely assembled states; and, for normalization $C_S + C_I + C_C = 1$. In principle, an exact solution to the equation can be obtained by finding the eigenvalues and eigenvectors of $Q$. However, this calculation requires a solution to a cubic equation and is too complicated to extract much significant physical information. Therefore, we first employed an approximate method that in our view clearly shows the dependence of the relevant parameters on the behavior of the signaling dynamics. The method can also be applied to other kinetic schemes. We also considered numerical solutions.

**Path summation of the Master equation**

Formally, we can compute $P(\tau_A < t)$ by considering a weighted sum over all paths that lead to the absorbing state, $K_A$.

$$P(\tau_A < t) = \sum_{steps, i=1}^{\infty} \sum_{j=1}^{3} C_j \sum_{branches, k=1} \omega\left(\left\{\sum_{jumps, l=1}^{i} \tau_l^{ijk}\right\} < t\right) \prod_{l=1}^{i} \prod_{m} \omega(\tau_l < \tau_{lm}) \quad (3)$$

The first summation decomposes $P(\tau_A < t)$ into separate contributions for each set of paths that contain equivalent numbers of steps required to reach the absorbing state, $K_A$; i.e. for i =1, all paths requiring one jump are considered, for i=2, all paths requiring two jumps are considered, etc. Since there can be more than one path containing i steps leading to $K_A$. The next summation considers the weighted probability that, a priori, a kinase is in one of three states: in solution (S), bound to a signaling incompetent complex (I), or bound to a signaling competent complex (C), i.e. $j \in \{S, I, C\}$ and for normalization, $C_S + C_I + C_C = 1$. We must then sum over each branch that are denoted with subscript k. A branch is defined here as a particular way in which a path of fixed i and j can be traversed. We account for the probability that a specific path, i, with j steps on the k[th] branch is taken by computing the probability $\omega\left(\left\{\sum_{jumps, l=1}^{i} \tau_l^{ijk}\right\} < t\right)$ that, in time $t$ an enzyme transitions through a given sequence of jumps, each of which involving a random waiting time $\tau_l^{ijk}$, leading to $K_A$ for the kth branch of the $j^{th}$ molecular state that takes i steps where a path of i steps is composed of individual steps, $l, 1 \leq l \leq i$. We then avoid over counting by taking the union of this probability with the joint probability, $\prod_{l=1}^{i} \prod_{m} \omega(\tau_l < \tau_{lm})$, that no transitions are made in the $l^{th}$ step along the path to any state, m, not along the considered path; $\tau_l$ is the waiting time to transition along the $l^{th}$ step of the path and $\tau_{lm}$ is the waiting time for a transition at the $l^{th}$ step to a position m that is not along the selected path. This term, $\prod_{l=1}^{i} \prod_{m} \omega(\tau_l < \tau_{lm})$, ensures that each transition $l \to l+1$ takes place before any transitions $l \to m$ to points not along the given path.

Such a path summation is difficult to compute exactly but conveniently lends itself to approximate evaluations. We first denote the contribution of each path requiring i steps, $a_i$ so that $P(\tau_A < t) = \sum_{i}^{\infty} a_i$. Then, the contribution of $a_i$ to the overall cumulative distribution, $P(\tau_A < t)$, decreases monotonically with increasing i. So, $a_i \geq a_{i+1}$ and thus

$$\sum_{j=1}^{3} C_j \sum_{branches, k=1} \omega\left(\left\{\sum_{jumps, l=1}^{i} \tau_l^{ijk}\right\} < t\right) \prod_{l=1}^{i} \prod_{m} \omega(\tau_l < \tau_m)$$
$$\geq \sum_{j=1}^{3} C_j \sum_{branches, k=1} \omega\left(\left\{\sum_{jumps, l=1}^{i+1} \tau_l^{(i+1)jk}\right\} < t\right) \prod_{l=1}^{i+1} \prod_{m} \omega(\tau_l < \tau_{lm})$$
(4)

Thus, the sum can be truncated at all paths requiring i steps with an error that is bounded by $O(a_{i+1})$. As the number of steps, i, increases, the total contribution of each path becomes smaller by a factor involving the ratio of the total contribution to $P(\tau_A < t)$ for the path containing i+1 and i steps; i.e.

$$\frac{a_{i+1}}{a_i} = \frac{\sum_{j=1}^{3} C_j \sum_{branches,k=1} \omega\left(\left\{\sum_{jumps,l=1}^{i+1} \tau_l^{(i+1)jk}\right\} < t\right) \prod_{l=1}^{i+1} \prod_m \omega(\tau_l < \tau_{lm})}{\sum_{j=1}^{3} C_j \sum_{branches,k=1} \omega\left(\left\{\sum_{jumps,l=1}^{i} \tau_l^{ijk}\right\} < t\right) \prod_{l=1}^{i} \prod_m \omega(\tau_l < \tau_{lm})} \quad (4)$$

We can simplify this formula by making use of two identities that hold for continuous time Markov chains. For two independent random variables, $\tau_i$ and $\tau_j$, that are exponentially distributed with time constants, $k_i$ and $k_j$, the probability of $\tau_i$ being less than $\tau_j$, $\omega(\tau_i < \tau_j)$ is $\omega(\tau_i < \tau_j) = \frac{k_i}{k_i + k_j}$. Also, for a sum of n exponentially distributed random variables $\tau_i; \tau_i \in \{\tau_1, \tau_2, ..., \tau_n\}$ with time constants, $k_i; k_i \in \{k_1, k_2, ..., k_n\}$, the probability of the sum of n independent random variables being less than $t$ is a convolution of those variables, $\omega\left(\sum_i^n \tau_i\right) < t) = \omega(\tau_1 < t) \otimes ... \otimes \omega(\tau_n < t)$ where the $\otimes$ symbol denotes a convolution, and has the following form:

$\omega\left(\sum_i^n \tau_i\right) < t) = 1 - \sum_{i=1}^{n}\left(\prod_{j \neq i} \frac{k_j}{k_j - k_i}\right) e^{-k_i t}$. By substituting these two expressions where appropriate,

$$\frac{a_{i+1}}{a_i} = \frac{\sum_{j=1}^{3} C_j \sum_{branches,k=1} \sum_{n=1}^{i+1}\left(1 - \prod_{j \neq n} \frac{k_j}{k_j - k_n} e^{-k_n t}\right) \prod_{l=1}^{i+1} \prod_m \frac{k_l}{k_l + k_m}}{\sum_{j=1}^{3} C_j \sum_{branches,k=1} \sum_{n=1}^{i}\left(1 - \prod_{j \neq n} \frac{k_j}{k_j - k_n} e^{-k_n t}\right) \prod_{l=1}^{i} \prod_m \frac{k_l}{k_l + k_m}} \quad (5)$$

This formula can be rearranged by factoring out the i+1 term inside the summation over the different "branches"

$$\frac{a_{i+1}}{a_i} = \frac{\sum_{j=1}^{3} C_j \sum_{branches,k=1} \left(\left(\left(\sum_{n=1}^{i} 1 - \beta_n \gamma_n\right) + \alpha\right)\kappa_{i+1}\right)}{\sum_{j=1}^{3} C_j \sum_{branches,k=1} \sum_{n=1}^{i}(1-\gamma_n)\kappa_i} \quad (6)$$

where, $\alpha = \prod_{j \neq i+1} \frac{k_j}{k_j - k_{i+1}} e^{-k_{i+1}t}$, $\beta_n = \frac{k_{i+1}}{k_{i+1} - k_n}$, $\gamma_n = \prod_{j \neq i+1} \frac{k_j}{k_j - k_n} e^{-k_n t}$, and

$\kappa_i = \prod_{l=1}^{i} \prod_{m} \frac{k_l}{k_l + k_{lm}}$. The error, E, that is introduced by truncating the sum at a given number of steps is

$$E = O\left(\frac{\sum_{j=1}^{3} C_j \sum_{branches,k=1} \left(\left(\left(\sum_{n=1}^{i} 1 - \beta_n \gamma_n\right) + \alpha\right)\kappa_{i+1}\right)}{\sum_{j=1}^{3} C_j \sum_{branches,k=1} \sum_{n=1}^{i}(1-\gamma_n)\kappa_i}\right) \quad (7)$$

From the formula in eq. 6, we see that many conditions allow for $\frac{a_{i+1}}{a_i} << 1$, in which case, the summation quickly decays and can be truncated at i steps. Moreover, any significant difference in time scales for processes in successive steps results in such a decrease.

Now we consider the application of this formalism to the model. Summing eq. 3 from j=1 to j=3 gives us

$$P(\tau_A < t) = P(\tau_{A,solution} < t) + P(\tau_{A,incomplete} < t) + P(\tau_{A,complete} < t). \quad (8)$$

This eq. 8 gives us the cumulative density as a composition of many terms contributing from initial states; from solution, incomplete complexes, and complete complexes where the sum is carried out up to $i = 3$ steps,

$$P(\tau_{A,solution} < t) = C_s \begin{cases} \left(P(\tau_7 < t)P(\tau_7 < \tau_1)P(\tau_7 < \tau_5)\right) + \\ \left(P(\tau_5 + \tau_8 < t)P(\tau_5 < \tau_7)P(\tau_5 < \tau_1)P(\tau_8 < \tau_4)P(\tau_8 < \tau_6)\right) + \\ \left(P(\tau_1 + \tau_2 + \tau_7 < t)P(\tau_1 < \tau_5)P(\tau_1 < \tau_7)P(\tau_2 < \tau_3)P(\tau_7 < \tau_1)P(\tau_7 < \tau_5)\right) + \\ \left(P(\tau_1 + \tau_3 + \tau_8 < t)P(\tau_1 < \tau_5)P(\tau_1 < \tau_7)P(\tau_3 < \tau_2)P(\tau_8 < \tau_4)P(\tau_8 < \tau_6)\right) \end{cases}$$

$$P(\tau_{A,incomplete} < t) = C_I \begin{cases} \left(P(\tau_2 + \tau_7 < t)P(\tau_2 < \tau_3)P(\tau_7 < \tau_1)P(\tau_7 < \tau_5)\right) + \\ \left(P(\tau_3 + \tau_8 < t)P(\tau_3 < \tau_2)P(\tau_8 < \tau_4)P(\tau_8 < \tau_6)\right) + \\ \left(P(\tau_2 + \tau_5 + \tau_8 < t)P(\tau_2 < \tau_3)P(\tau_5 < \tau_1)P(\tau_5 < \tau_7)P(\tau_8 < \tau_4)P(\tau_8 < \tau_6)\right) + \\ \left(P(\tau_3 + \tau_6 + \tau_7 < t)P(\tau_3 < \tau_2)P(\tau_6 < \tau_4)P(\tau_6 < \tau_8)P(\tau_7 < \tau_1)P(\tau_7 < \tau_5)\right) \end{cases}$$

.

$$P(\tau_{A,complete} < t) = C_C \begin{cases} P(\tau_8 < t)P(\tau_8 < \tau_4)P(\tau_8 < \tau_6) + \\ P(\tau_6 + \tau_7 < t)P(\tau_6 < \tau_4)P(\tau_6 < \tau_8)P(\tau_7 < \tau_1)P(\tau_7 < \tau_5) + \\ \left(P(\tau_3 + \tau_4 + \tau_8 < t)P(\tau_4 < \tau_6)P(\tau_4 < \tau_8)P(\tau_3 < \tau_2)P(\tau_8 < \tau_4)P(\tau_8 < \tau_6)\right) + \\ \left(P(\tau_2 + \tau_4 + \tau_7 < t)P(\tau_4 < \tau_6)P(\tau_4 < \tau_8)P(\tau_2 < \tau_3)P(\tau_7 < \tau_1)P(\tau_7 < \tau_5)\right) \end{cases}$$

(9)

Fig. 2 considers eq. 9 as compared to the exact numerical solution of eq. 1. In each case, the survival probability $S(t) = 1 - P(\tau_A < t)$ is plotted. In Fig. 2a, processes involving activation, $\tau_7$ and $\tau_8$, dominate the kinetic pathway and the approximate solution is quantitatively accurate at all times. In Figs. 2b,c, $\tau_7$ and $\tau_8$ are still dominate time scales but other transitions within the pathway also contribute significantly; as a result, eq. 9 still captures much of the qualitative behavior such as the shapes and overall time to relaxation but quantitative deviations in eq. 9 and eq. 2 are readily apparent. Lastly in Fig. 2d, each state transition contributes to the arrival at the absorbing state. In this scenario, eq. 9 provides a poor description of the dynamics in eq. 2. Combined, the plots in Fig. 2 illustrate the regimes of validity of the use of such a path summation technique in extracting useful physical information from the kinetic model in eq. 2.

**Application to the dependence on scaffold density**

Now, consider the case in which scaffold proteins are present in negligible amounts. In this case, a separation of time scales is apparent with one dominant time scale controlling kinase activation. For low scaffold concentrations ($C_S \sim 1$), most kinases are present in solution and the transition to either scaffold-bound state is very slow, i.e. $k_7 >> k_5$ and $k_7 >> k_1$, the summation can therefore, with small error, be cut off at one step giving:

$$P(\tau_A < t) = 1 - \left(\frac{k_7}{k_7 + k_1}\right)\left(\frac{k_7}{k_7 + k_5}\right) e^{-k_7 t} + h.o. \approx 1 - e^{-k_7 t}. \qquad (10)$$

If signaling is only allowed to take part on a scaffold (i.e. $k_7 = 0$) then by the same argument,

$$P(\tau_A < t) \approx 1 - e^{-k_5 t}. \qquad (11)$$

Now consider the situation where scaffold concentration is very high (i.e. $C_I \approx 1$; in this case, we consider paths starting from $K_I$ that end in the active state. Starting at $K_I$, there are two branches that lead to $K_A$, $K_I \xrightarrow{2} K_S \xrightarrow{7} K_A$ and $K_I \xrightarrow{3} K_C \xrightarrow{8} K_A$. However, the dominant time scale in this scenario is $k_3^{-1}$. Keeping only the branch that involves process 3, since $k_4 >> k_3$ and $k_8 >> k_3$ we have

$$P(\tau_A < t) = P(\tau_3 + \tau_8 < t)P(\tau_3 < \tau_4)P(\tau_8 < \tau_4)P(\tau_8 < \tau_6)$$

$$= 1 - \left(\frac{k_3}{k_3 + k_4}\right)\left(\frac{k_8}{k_8 + k_4}\right)\left(\frac{k_8}{k_8 + k_6}\right)\left(\frac{k_8}{k_8 - k_3}e^{-k_3 t} + \frac{k_3}{k_3 - k_8}e^{-k_8 t}\right) + h.o. \quad (12)$$

$$\approx 1 - \left(\frac{k_3 k_8^2}{k_4(k_8 + k_4)(k_8 + k_6)}\right)e^{-k_3 t}$$

In the intermediate regime of scaffold density, no such separation of time scales is apparent; also, there is significant probability that a kinase resides in any of the three initial states. Typical equilibrium disassociation constants, $K_d$, for association to a scaffold are $\sim 1\mu M$ (1kT/molecule $\sim$ 0.6 kCal/mol) and this corresponds to a binding energy $F = -k_b T \ln(K_d)$ of 12-15kT(12). Given this binding affinity and typical kinase concentrations that result in an excess number of signaling residing in solution than potentially bound to a scaffold(13), ~85% ($P_{scaf} \approx 0.85$) of proteins are bound to a scaffold. This implies that $(P_{scaf})^n$, for n =3 binding sites, ~72% of the bound signaling proteins exist in fully assembled signaling-competent complexes. In this situation, all kinetic pathways are important and the summation in eq. 3 requires more terms than the truncation at i =3 that is contained in eq. 9. This expression gives us, when decomposed into a contribution from solution, incomplete complexes, and complete complexes, a superposition of many exponential terms. In this case, one can see from the formula that the cumulative distribution is a composition of many characteristic time scales that govern signaling dynamics. We show this by solving eq. 2 numerically in Fig 3.

Conceptually, from the physical processes occurring in our model, there are multiple ways in which parameters are affected by changes in scaffold concentration. One effect is the alteration of $C_I$ and $C_S$, the number of proteins that exist in complexes that are incomplete and signaling competent respectively. The rates of transitions between states corresponding to processes occurring with times $\tau_1, \tau_3$, and $\tau_5$ are each affected by the relative scaffold concentrations and thus $C_I$ and $C_S$. We consider a simple parameterization in which the rates of the transitions to each scaffold containing state are proportional to the initial concentrations of that state. In this scheme, $k_1 = k_1^0 C_I$, $k_3 = k_3^0 C_S$, and $k_5 = k_5^0 C_S$. With this parameterization, Fig. 3a contains plots of $S(t)$, obtained from the numerical solution of eq. 2. For unbinding transitions, a typical off rate(12) ($\sim 0.1 s^{-1}$) is used. Also, the rate of activation of a kinase on a completely assembled is taken to be ~100 times greater than that in solution.

As seen in Fig. 3a, when kinase activation is primarily occurring on the scaffold (solid line), $S(t)$ sharply decays. When a significant population of each state is present (dash-dotted line), the decay of $S(t)$ and thus signaling occurs smoothly across multiple decades. In the other two cases, proteins are either confined to incomplete signaling complexes (dotted line) and primarily in solution (dashed line) activation occurs via a two-stage mechanism. The first stage involves the kinases that transition to the fully, assembled ($K_C$) state and are quickly activated. The second stage of activation involves the population of kinases that are slowly activated in the solution state.

Alternatively, it is conceivable that the rates, $k_1$, $k_3$, and $k_5$, depend on scaffold concentration in a more complicated manner. In Fig. 3b, we considered the rates to depend on scaffold concentration in a nonlinear, quadratic manner; i.e. $k_1 = k_1^0 (C_I)^2$,

$k_3 = k_3^0 (C_S)^2$, and $k_5 = k_5^0 (C_S)^2$. The primary effect as seen in the plots is to broaden the shape of the decay curves.

We showed that a multi-state kinetic model with Markovian dynamics can give rise to complicated multi-exponential kinetics in some range of scaffold concentrations. It is likely, however, that transitions between these states in actuality have more complicated transitions. The minimal model presented, therefore, illustrates the simplest mechanism that illustrates how a waiting time distribution for signal transduction can be affected by scaffold concentration.

**Summary**

We showed, in a simple model, the competition of the many processes that govern scaffold mediated signal transduction. We applied an approximate technique, involving a weighted path summation, along with exact numerical solutions to show how the shape of waiting time distribution of kinase activation and thus the nature of the signal output can depend on scaffold density within the framework of a minimal kinetic model. The shape of such a distribution is believed to be important in detecting the time scale dependence of biochemical signaling(14, 15). Also, since such models occur in many difference areas of biology and chemistry, it may be interesting to investigate the behaviors of other models(16, 17) in the context of this weighted path summation technique.

**Acknowledgements**

This work is supported in part from an NIH PO1AI071195-01 and an NIH Director's Pioneer Award granted to Arup Chakraborty. I thank Arup Chakraborty for his support. I am also grateful to Fei Liang and Roger York for helpful discussions.

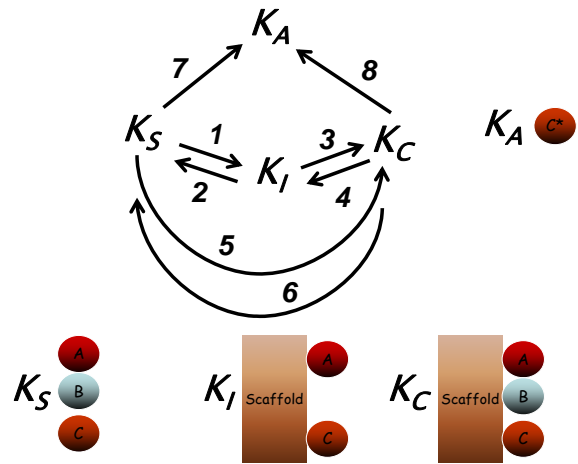

Figure 1.  **A three state kinetic mechanism that characterizes scaffold mediated cell signaling**

Important time scales in scaffold mediated signaling shown in a graph of a multi-state kinetic model whose dynamics are governed by 8 transitions. The final kinase molecule in a signaling cascade (C) can transition between three states denoted with subscripts: in solution ($K_S$), bound to a signaling competent complex ($K_C$), bound to a signaling incompetent complex ($K_I$), and eventually reaches an activated absorbing state, ($K_A$), in which $C \to C^*$ where * denotes the activated form in the figure. The relative amount of each state and the rates of transitioning are functions of the density of scaffold proteins

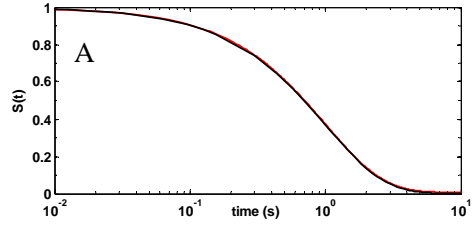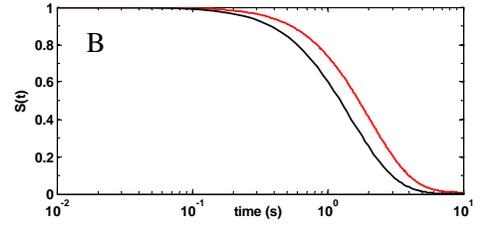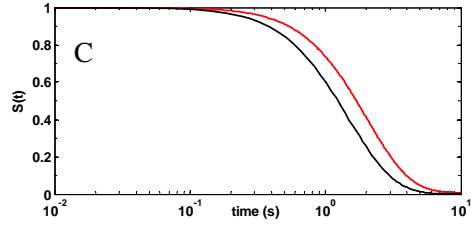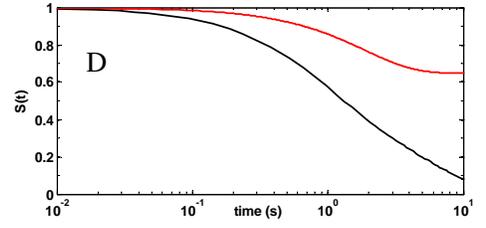

Figure 2. **Comparison of the path summation approximation and the exact numerical solution of the three-state model.**

Exact numerical solutions of eq. 1 are compared to eq. 9 which constitutes a truncation of the path summation in eq. 4 at i=3 steps. Exact solutions are shown in solid lines and the approximate solutions in dotted lines. Parameters used are: a.) $k_1 = 0.001$, $k_2 = 0.002$, $k_3 = 0.003$, $k_4 = 0.004$, $k_5 = 0.005$, $k_6 = 0.006$, $k_7 = k_8 = 1.0$, $C_s = 0.5$, $C_C = 0.5$, $C_I = 0.0$. b.) $k_1 = 0.001$, $k_2 = 1.002$, $k_3 = 1.003$, $k_4 = 0.004$, $k_5 = 0.005$, $k_6 = 0.006$, $k_7 = k_8 = 1.0$, $C_s = 0.0$, $C_C = 0.0$, $C_I = 1.0$. c.) $k_1 = 0.101$, $k_2 = 0.102$, $k_3 = 0.103$, $k_4 = 0.104$, $k_5 = 0.105$, $k_6 = 0.106$, $k_7 = k_8 = 1.0$, $C_s = 0.5$, $C_C = 0.5$, $C_I = 0.0$. d.) $k_1 = 1.001$, $k_2 = 1.002$, $k_3 = 1.003$, $k_4 = 1.004$, $k_5 = 1.005$, $k_6 = 1.006$, $k_7 = k_8 = 1.0$, $C_s = 0.333$, $C_C = 0.333$, $C_I = 0.333$.

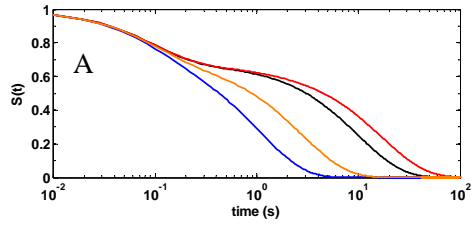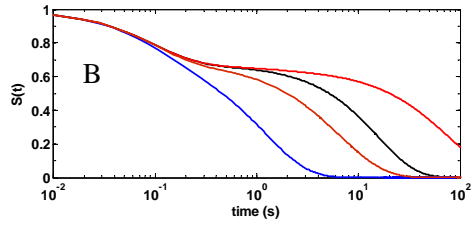

Figure 3. **Variations in scaffold concentration**

Numerical solutions of eq. 1 are considered with base parameters: $k_1^0 = 1.001$, $k_2 = 0.102$, $k_3^0 = 1.003$, $k_4 = 0.104$, $k_5^0 = 1.005$, $k_6 = 0.106$, $k_7 = 0.1$ $k_8 = 10.0$. $C_C$, $C_I$, and $C_s = 1 - C_C - C_I$ are varied. a.) linear ($C_i = k_i^0 C_j$) and b.) quadratic ($C_i = k_i^0 (C_j)^2$, (for i=1,3,5 and j =S,I), parameterizations of the effects of the rates by changes in scaffold concentration are used. Four cases are considered: $C_C = 0.05$, $C_I = 0.05$, $C_s = 0.90$ (dashed lines) ; $C_C = 0.90$, $C_I = 0.05$, $C_s = 0.05$ (solid lines) ; $C_C = 0.05$, $C_I = 0.90$, $C_s = 0.05$ (dotted lines) ; $C_C = 0.333$, $C_I = 0.333$, $C_s = 0.333$ (dash-dotted lines).